\documentclass[aps,preprint]{revtex4}
\usepackage{graphicx}
\begin{document}
\title{Calculation of the staggered spin correlation in the framework of the Dyson-Schwinger approach}
\author{Jian-Feng Li$^{1,2}$, Hong-Tao Feng$^{3,2}$, Yu Jiang $^{4,2}$,  Wei-Min Sun$^{5,6,2}$, Hong-Shi Zong$^{5,6,2}$}
\address{$^{1}$ College of  Mathematics and Physics, Nantong University, Nantong 226019, China}
\address{$^{2}$ State Key Laboratory of Theoretical Physics, Institute of Theoretical Physics, CAS, Beijing 100190, China}
\address{$^{3}$Department of Physics, Southeast University, Nanjing, China}
\address{$^{4}$ College of Mathematics, Physics and Information Engineering, Zhejiang Normal University, Jinhua 321004, China}
\address{$^{5}$ Department of Physics, Nanjing University, Nanjing 210093, China}
\address{$^{6}$ Joint Center for Particle, Nuclear Physics and Cosmology, Nanjing 210093,
China}

\begin{abstract}

Based on the linear response of the fermion propagator with respect
to an external field, we first derive a model-independent expression
for the staggered spin susceptibility in which the influence of the
full pseudoscalar vertex function is included. This expression for
the staggered spin susceptibility is quite different from that given
in the previous literature. The numerical values of the staggered
spin susceptibility are calculated within the framework of the
Dyson-Schwinger approach. Our numerical result shows that the
nonperturbative dressing effects on  the fermion propagator is very
important when one studies the staggered spin susceptibility which
corresponding to antiferromagnetic correlation in both Nambu phase
and Winger phase.



Email: zonghs@chenwang.nju.edu.cn

\end{abstract}

\maketitle Quantum electrodynamics in (2+1) dimensions (QED$_3$) has
attracted much interest over the past few years. It has many
features similar to QCD, such as spontaneous chiral symmetry
breaking in the chiral limit and confinement
\cite{a1,a2,a3,a4,a5,a6,feng,a7,a8,a9,a10,a11,a12}. Moreover, it is
super-renormalizable, so it does not suffer from the ultraviolet
divergence which are present in QED$_4$. Due to these reasons it can
serve as a toy model of QCD. In parallel with its relevance as a
tool through which to develop insight into aspects of QCD, QED$_3$
is also found to be equivalent to the low-energy effective theories
of strongly correlated electronic systems. Recently, QED$_3$ has
been widely studied in high T$_c$ cuprate superconductors
\cite{a13,a14,a15,a16,a17,a18,a19,a20,a21,a22} and graphene
\cite{a23,a24,a25}.

Dynamical chiral symmetry breaking (DCSB) occurs when the massless
fermion acquires a nonzero mass through nonperturbative effects at
low energy, but the Lagrangian keeps chiral symmetry when the
fermion mass is neglected. In a four-fermion interaction model
 Nambu and Jona-Lasinio first adopted the mechanism of
DCSB to generate a nonzero mass for the fermion from nothing solely
through interactions \cite{a26}. The Dyson-Schwinger equations
(DSEs) provide a natural framework within which to explore DCSB and
related phenomena. In 1988, T. Appelquist et al. \cite{a4} studied
DCSB in massless QED$_3$ with $N$ fermion flavors by solving the
DSEs  for fermion self-energy in leading order of the $1/N$
expansion and found DCSB occurs when $N$ is less than a critical
number $N_c$. Later D. Nash showed that the gauge-invariant critical
number of fermion flavor still exists by considering higher order
corrections to the gap equation \cite{a5}. In 1995, P. Maris solved
the coupled DSEs with a set of simplified vertex functions and
obtained the critical number of fermion flavor $N_c$=3.3
\cite{a6,a7}. Recently, in massless unquenched $\rm{QED_{3}}$,
Fischer et al. \cite{a11} self-consistently solved a set of coupled
DSEs and obtained $N_{c}^{crit}\approx 4$ by using more
sophisticated vertex ansatz in unquench $\rm{QED_{3}}$.

From the differences in the temperature dependence of the Cu and O
site relaxation rates in NMR experiment, it is shown that there
exist antiferromagnetic correlations in the underdoped cuprates.
However, within the slave boson mean field theory of the t-J model,
the staggered spin correlation tends to zero in the infrared region.
That is to say, the mean field treatments lose a lot of
antiferromagnetic correlation. In order to explain the above puzzle,
Wen et al. considered the U(1) gauge fluctuations corrections to the
staggered spin susceptibility at order 1/N in the Algebraic spin
liquid (ASL) based on slave boson treatment of t-J model and found
that the dressed fermion propagator acquires an anomalous dimension
exponent compared with the free spinon propagator in the chiral
symmetric phase (Wigner phase), which enhances the staggered spin
correlation and restores the antiferromagnetic order which is
lost at the mean-field level \cite{a18,a19,a20,a21}. It is clear
that the staggered spin susceptibility plays a crucial role in ASL
where the anomalous dimension exponent indicates the non-Fermi
liquid behavior in the pseudogap phase in effective $\rm{QED_{3}}$ theory
of cuprate superconductor.

On the other hand, based on the phase fluctuation model \cite{a27},
Franz et al. proposed a new quantum liquid-the Algebraic Fermi liquid
(AFL) to describe the pseudogap state in the effective
$\rm{QED_{3}}$ theory \cite{a28,a29}. By studying the
gauge-invariant response function in AFL, Franz et al. also found
that the full staggered spin susceptibility exhibits a nontrivial
anomalous dimension exponent to 1/N order in Wigner phase. This
result is in agreement with the finding of Refs. \cite{a20,a21}.
However, Liu et al. argue that once the fermion acquires a constant
mass in DCSB phase (Nambu phase), the spin staggered correlation
function defined in Ref. \cite{a21} is nonzero even at the mean
field level and antiferromagnetic order gets restored \cite{a30}.

Up to now, in all the above literatures, the theoretical calculations of
the spin susceptibility are usually done in the framework of
perturbation theory where only the leading 1/N order corrections to the
staggered spin correlation function is added to the mean field
level. In this paper, we will study the staggered spin susceptibility
using a nonperturbative method since the appearance of antiferromagnetic
order is a nonperturbative phenomenon.

Our starting point is the Lagrangian density of massless effective
QED$_3$ with N-flavored Dirac fermion
\begin{equation}\label{eq1}
\mathcal{L}=\sum^N_{i=1}\bar\psi_i\gamma_{\mu}(\partial_{\mu}+\mathrm{i}e
A_{\mu})\psi_i+\frac{1}{4}F^2_{\mu\nu}.
\end{equation}
In this context, we are focusing on the case of N=2 since the nature
of the effective $\rm{QED_{3}}$ theory at N = 2 is most interesting for
true electronic systems. In $\rm{QED_{3}}$, the dimensional
coupling constant $\alpha=e^{2}$, which provides an intrinsic mass
scale similar to $\Lambda_{QCD}$ in QCD.  For simplicity, we set
$\alpha=1$ in this paper. The $4 \times 4$ gamma matrices can be
defined as $\gamma_{\mu}=\sigma_{3}\otimes(\sigma_{2},\sigma_{1},-\sigma_{3})$,
which satisfy the Clifford algebra
$\{\gamma_{\mu},\gamma_{\nu}\}=2\delta_{\mu\nu} (\mu,\nu=0,1,2)$. It
is well known that the DSEs provides a successful description of
various nonperturbative aspects of strong interaction physics in QCD
\cite{DSER1,DSER2}. We naturally expect that it could be a useful
nonperturbative approach in the study of spin susceptibilities and in this paper we will give a full formula for the staggered spin susceptibility. 

In this work, we shall employ the linear response
theory of fermion propagator to study the staggered spin correlation
function \cite{a28,a29}. Our starting point is the usual $\rm{QED_{3}}$
Lagrangian added with an additional coupling term $\Delta\mathcal
{L}=\bar{\psi}_{\gamma}(x)(\gamma_{5})_{\gamma\delta}\psi_{\delta}(x)\mathcal
{V}(x)$, where $\gamma_{5}=\sigma_{2}\otimes\textbf{1}$ and
$\mathcal{V}(x)$ is a variable external pseudoscalar field. The fermion
propagator $\mathcal{G}_{\alpha\beta}[\mathcal {V}](x)$ in the
presence of the external field $\mathcal{V}$ can be written as \cite{Zong-Vector,Zong-Axial,Zong-Tensor}
\begin{equation}\label{1}
    \mathcal{G}_{\alpha\beta}[\mathcal {V}](x)=\int\mathcal {D}\bar{\psi}\mathcal {D}\psi\mathcal
    {D}A\psi_{\alpha}(x)\bar{\psi}_{\beta}(0)exp\{-\int d^{3}x[\mathcal {L}+\Delta\mathcal
    {L}]\},
\end{equation}
where the subscripts denote the spinor indices. If we only consider
the linear response term of the fermion propagator
$\mathcal{G}_{\alpha\beta}[\mathcal {V}](x)$, we obtain
\begin{eqnarray}\label{2}
\mathcal{G}_{\alpha\beta}[\mathcal {V}](x) &=& \int\mathcal
{D}\bar{\psi}\mathcal {D}\psi\mathcal
{D}A\psi_{\alpha}(x)\bar{\psi}_{\beta}(0)exp\{-S[\bar{\psi},\psi,A]\}\nonumber\\
&&+\int\mathcal {D}\bar{\psi}\mathcal {D}\psi\mathcal {D}A\int
d^{3}y[\psi_{\alpha}(x)\bar{\psi}_{\beta}(0)\bar{\psi}_{\gamma}(y)(\gamma_{5})_{\gamma\delta}\psi_{\delta}(y)\mathcal
{V}(y)]exp\{-S[\bar{\psi},\psi,A]\}+\cdot\cdot\cdot,
\nonumber\\
 &=&G_{\alpha\beta}(x)+\mathcal {G}{^{\mathcal{V}}}_{\alpha\beta}(x)+\cdot\cdot\cdot
\end{eqnarray}
 where  $G_{\alpha\beta}(x)=\langle0\mid
T[\psi_{\alpha}(x)\bar{\psi}_{\beta}(0)]\mid 0\rangle$ is the
fermion propagator in the absence of the external field
$\mathcal{V}$, the linear response term of the fermion propagator
\begin{eqnarray}\label{2}
\mathcal {G}{^{\mathcal {V}}}_{\alpha\beta}(x)=\int d^{3}z
\langle0\mid
T[\psi_{\alpha}(x)\bar{\psi}_{\beta}(0)\bar{\psi}_{\gamma}(z)(\gamma_{5})_{\gamma\delta}\psi_{\delta}(z)]\mid
0\rangle \mathcal {V}(z)
\end{eqnarray}
and the ellipsis represents terms of higher order in $\mathcal{V}$.

Now we expand the inverse fermion propagator $\mathcal{G}^{-1}[\mathcal {V}]$ in powers of $\mathcal{V}$ as follows
\begin{equation}\label{2}
    \mathcal {G}^{-1}[\mathcal {V}]=\mathcal {G}^{-1}[\mathcal
    {V}]\mid _{\mathcal {V}=0} +\frac{\delta \mathcal {G}^{-1}[\mathcal {V}]}{\delta\mathcal
    {V}} \mid _{\mathcal {V}=0} \mathcal {V}+\cdot\cdot\cdot=G^{-1}+\mathcal
    {V}\Gamma_{P}+\cdot\cdot\cdot,
\end{equation}
which leads to the following formal expansion
\begin{equation}\label{2}
\mathcal {G}[\mathcal {V}]=G-G{\mathcal
    {V}}\Gamma_{P} G+\cdot\cdot\cdot.
\end{equation}
Here, the pseudoscalar vertex $\Gamma_{P}$ is defined as
\begin{equation}\label{3}
\Gamma_{P}(y_{1},y_{2},z)=\frac{\delta \mathcal {G}^{-1}{[\mathcal
{V}]}(y_{1},y_{2})}{\delta\mathcal
    {V}(z)} \mid _{\mathcal {V}=0}.
\end{equation}
Note that Eq. (7) is a compact notation and its explicit form reads
{\small\begin{eqnarray}\label{2} &&
\mathcal{G}_{\alpha\beta}[\mathcal
    {V}](x)=G_{\alpha\beta}(x)-\int d^{3}y_{1} d^{3}y_{2}
    d^{3}zG_{\alpha\gamma}(x-y_{1})[\Gamma_{P}(y_{1},y_{2},z)\mathcal
    {V}(z)]_{\gamma\delta}G_{\delta\beta}(y_{2}) +\cdot\cdot\cdot\\
    &=&G_{\alpha\beta}(x)-\int d^{3}z \int \frac{d^{3}P}{(2\pi)^{3}}\int
    \frac{d^{3}q}{(2\pi)^{3}}e^{-i(q+\frac{P}{2})x}e^{iP\cdot Z}G_{\alpha\gamma}(q+\frac{P}{2})[\Gamma_{P}(q,P)\mathcal
    {V}(z)]_{\gamma\delta} G_{\delta\beta}(q-\frac{P}{2})+\cdot\cdot\cdot.\nonumber
\end{eqnarray}}
Setting $x=0$  in Eq. (5) and comparing the linear response term in
Eq. (9), we obtain
\begin{eqnarray}\label{10}
&&\langle0\mid T[{{\psi}_{\alpha}}(0)\bar{\psi}_{\beta}(0)\bar{\psi}_{\gamma}(z)(\gamma_{5})_{\gamma\delta}\psi_{\delta}(z)]\mid 0\rangle\nonumber\\
&&=-\int \frac{d^{3}P}{(2\pi)^{3}}\int
    \frac{d^{3}q}{(2\pi)^{3}}e^{iP\cdot z}G_{\alpha\gamma}(q+\frac{P}{2})[\Gamma_{P}(q,P)]_{\gamma\delta} G_{\delta\beta}(q-\frac{P}{2}).
\end{eqnarray}
After multiplying $(\gamma_{5})_{\beta\alpha}$ on both sides of Eq.
(10) and summing over the spinor indices, we finally obtain
\begin{eqnarray}\label{10}
&&\langle0\mid T[\bar{\psi}_{\beta}(0)(\gamma_{5})_{\beta\alpha}{\psi}_{\alpha}(0)\bar{\psi}_{\gamma}(z)(\gamma_{5})_{\gamma\delta}\psi_{\delta}(z)]\mid 0\rangle\nonumber\\
&&=\int\frac{d^{3}P}{(2\pi)^{3}}\int\frac{d^{3}q}{(2\pi)^{3}}e^{iP\cdot
z}Tr[G(q+\frac{P}{2})[\Gamma_{P}(P,q)] G(q-\frac{P}{2})\gamma_{5}].
\end{eqnarray}

The staggered spin correlation function in momentum space is
\begin{equation}\label{3}
\langle S^{z}(x)S^{z}(0)\rangle
=\int\frac{d^{3}P}{(2\pi)^{3}}e^{iPx}\langle
S^{z}(P)S^{z}(-P)\rangle,
\end{equation}
where the z-component of the electron spin density operator
$S^{z}(x)=\bar{\psi}(x)\gamma_{5} \psi(x)$ \cite{a29}. From Eq. (11)
and Eq. (12), one obtain the following model-independent expression
for the staggered spin susceptibility
\begin{equation}\label{3}
\langle S^{z}(P)S^{z}(-P)\rangle
=\int\frac{d^{3}q}{(2\pi)^{3}}Tr[G(q+\frac{P}{2})\Gamma_{P}(q,P)
G(q-\frac{P}{2})\gamma_{5}].
\end{equation}
It should be noted that Eq. (13) is quite different from that given
in the previous literature \cite{a20,a21,a28,a29}. It reduces into
the staggered spin correlation defined in Ref. \cite{a29} when one
uses the bare pseudoscalar vertex $\Gamma_{P0}=\gamma_{5}$.

From Eq. (13) it can be seen that the staggered spin susceptibility
is closely related to the full fermion propagator and the
pseudoscalar vertex. Once the full fermion propagator and the
pseudoscalar vertex are known, one can calculate exactly the
staggered spin susceptibility. However, at present it is still not
possible to calculate the full fermion propagator and the
pseudoscalar vertex from the first principles of $\rm{QED_{3}}$. So,
in this work we will adopt the DSE approach to calculate the fermion propagator and the pseudoscalar vertex.

The general form of the dressed fermion propagator is:
\begin{equation}\label{eq2}
G^{-1}(p)=i\gamma\cdot pA(p^{2}) +B(p^{2}),
\end{equation}
where $A(p^{2})$ is the wave function renormalization and $B(p^{2})$
is the self-energy function. As usual, one calls the phase in which
$B(p^{2})=0$ the Wigner phase and the phase in which $B(p^{2})\neq0$
the Nambu phase. Under rainbow approximation, the DSEs for the fermion
propagator can be written as:
\begin{equation}\label{3}
G^{-1}(p^{2})=i\gamma\cdot p+\Sigma (p^{2})=i\gamma\cdot
p+\int\frac{d^{3}q}{(2\pi)^{3}}D_{\mu\nu}(p-q)\gamma_{\mu}G(q)\gamma_{\nu},
\end{equation}
where the gauge boson  propagator in Landau gauge reads
\begin{equation}\label{3}
D_{\mu\nu}(q)=\frac{1}{q^{2}[1+\Pi(q^{2})]}(\delta_{\mu\nu}-\frac{q_{\mu}q_{\nu}}{q^{2}})
\end{equation}
with $\Pi(q^{2})$ being the polarization function. The pseudoscalar vertex $\Gamma_{P}$
satisfies an inhomogeneous Bethe-Salpeter equation, which in the
ladder approximation reads
\begin{equation}\label{3}
\Gamma_{P}(q,P)=\gamma_{5}-\int\frac{d^{3}k}{(2\pi)^{3}}\gamma_{\mu}D_{\mu\nu}(k-q)G(k+\frac{P}{2})\Gamma_{P}(k,P)G(k-\frac{P}{2})\gamma_{\nu}.
\end{equation}
We now focus on the low-energy behavior of the staggered spin
susceptibility in Nambu phase and Wigner phase. From Lorentz structure analysis,
$\Gamma_{P}(q)$ can be written as:
\begin{equation}
\Gamma_{P}(q)=\gamma_{5}F(q^{2})+i\gamma\cdot q\gamma_{5} H(q^{2}).
\end{equation}
Substituting Eq. (13) and Eq. (17) into Eq. (12), we finally obtain the staggered spin susceptibility in low energy
limit
\begin{equation}
 \langle S^{z}(0)S^{z}(0)\rangle=\int\frac{d^{3}q}{(2\pi)^{3}}\frac{4F(q^{2})}{[q^{2}A^{2}(q^{2})+B^{2}(p^{2})]},
\end{equation}
where
\begin{equation}
F(q^{2})=1+2\int\frac{d^{3}k}{(2\pi)^{3}}\frac{F(k^{2})}{(k-q)^{2}(1+\Pi[(k-q)^{2}])[k^{2}A^{2}(k^{2})+B^{2}(k^{2})]}.
\end{equation}
From Eq. (18) we see that
once the functions $A(p^2)$, $B(p^2)$ and $F(p^2)$ are known, the staggered spin susceptibility can be calculated. The remaining task is then to calculate
these functions by numerically solving the corresponding DSEs.

By numerically solving the coupled DSEs for the fermion propagator, we can obtain the wave-function renormalization $A(p^{2})$ and the fermion
self-energy function $B(p^{2})$. From these we can study the staggered spin susceptibility. In this paper, following Refs. \cite{a6,feng}, we choose the vertex ansatz
$\Gamma_\nu(p,k)=\frac{1}{2}[A(p^2)+A(k^2)]\gamma_\nu$ (the BC$_{1}$
vertex). Thus in the Landau gauge the coupled DSEs reads
\begin{equation}\label{eq10}
A(p^{2})=1+\frac{1}{p^{2}}\int\frac{\mathrm{d}^3k}{(2\pi)^3}\frac{A(p^{2})+A(k^{2})}{A^{2}(k^{2})k^{2}+B(p^{2})}\frac{A(k^{2})(p\cdot
q)(k\cdot q)/q^{2} }{[q^{2}(1+\Pi(q^{2}))]},
\end{equation}
\begin{equation}\label{eq11}
B(p^2)=\int\frac{\mathrm{d}^3k}{(2\pi)^3}\frac{B(k^2)[A(p^2)+A(k^2)]}{[A^2(k^2)k^2+B^2(k^2)][q^2(1+\Pi(q^2))]},
\end{equation}
\begin{equation}\label{eq12}
\Pi(q^2)=N\int
\frac{\mathrm{d}^3k}{(2\pi)^3}\frac{A(k^2)A(p^2)[A({p}^2)+A(k^2)]}{q^2[A^2(k^2)k^2+B^2(k^2)]}
\frac{[2k^2-4k\cdot q-6(k\cdot q)^2/q^2]}{[A^2(p^2)p^2+B^2(p^2)]},
\end{equation}
where $q=p-k$. By numerically solving the above coupled DSEs, one can obtain the momentum dependence of $A(p^2)$ (in both Nambu phase and Wigner phase) and $B(p^2)$ (in Nambu phase), which is plotted in Fig. 1 and Fig. 2, respectively.
\begin{figure}[htbp]
\includegraphics[width=12cm]{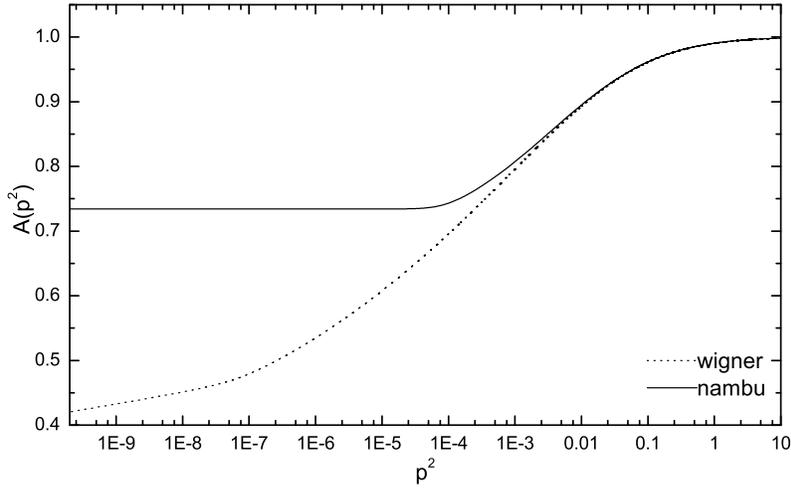}
\caption{The momentum dependence of $A(p^2)$ with $N=2$ in both Nambu phase and Wigner phase}  \label{Fig.1}
\end{figure}
\begin{figure}[htbp]
\includegraphics[width=12cm]{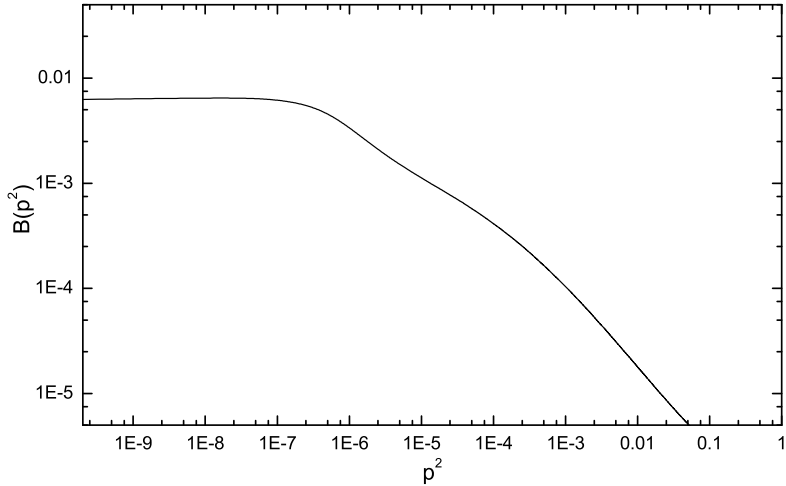}
\caption{The momentum dependence of $B(p^2)$ with $N=2$ in Nambu phase}  \label{Fig.2}
\end{figure}
From Fig.1 it can be seen that $A(p^2)$ in Wigner phase has a power law behavior
in the infrared region
\begin{equation}
A(p^2)=c p^{2\kappa},
\end{equation} 
where the constant $c$ and the power $\kappa$ agree with the result in Ref. \cite{a11}.  Here, we would like to stress that the
characteristic anomalous exponents of stressed fermion propagator of
the ASL or AFL have been widely studied in pure $\rm{QED_{3}}$
model. It has been conjectured long ago  that the vector dressing
function $A(p^{2})$ in Landau gauge is given by power laws in the
infrared region in the chiral symmetric phase of massless
$\rm{QED_{3}}$ \cite{a31}. But this conjecture has been validated
only recently both analytically and numerically in Ref. \cite{a11}
for the $1/N_{f}$ and more complete truncation schemes. The authors
of Ref. \cite{a11} obtained the explicit anomalous dimension of the
fermion vector dressing function in the infrared region by
numerically solving the full coupled set of DSEs. They found that
the corresponding anomalous dimension in Wigner phase is different
from the old $1/N_{f}$ expression given in Ref. \cite{a31}. In the
DCSB phase the vector dressing function $A(p^2)$ does not have a
power law behavior in the infrared region. The reason for this is as
follows. In the DCSB phase a second scale, i.e., the dynamically
generated fermion mass, emerges. This mass enters the photon polarisation and
the equation for the $A(p^{2})$ function and prevents a power law in
the infrared. The deeper reason is that power laws always correspond
to some kind of conformality, which usually does not exist once
there exists a scale (the fermion mass) in the system. As discussed
in Ref. \cite{a29}, this change of behavior reflects the difference
in physics between the AFL and the AF phase and should lead to a
difference between the staggered spin susceptibility in the chiral
symmetric phase and that in the DCSB phase, independent of the
regularization procedure. In this paper, we try to show that the
staggered spin susceptibility in  Wigner phase is different from
that in  Nambu phase, since the anomalous dimension emerges
and plays an important role only in Wigner phase.

Before numerically calculating the staggered spin susceptibility, let us analyze the large momentum behavior of the integrand in Eq. (18). From the large momentum behavior of $A(p^2)$, $B(p^2)$ and $F(p^2)$ we see
that the staggered spin susceptibility given by Eq. (18) is
linearly divergent and this divergence cannot be eliminated through the standard renormalization procedure. This is very similar to the case of chiral susceptibility in QCD, which is quadratically divergent  \cite{Aoki, He}. In order to extract something meaningful from the staggered spin correlation, one needs to subtract the linear divergence of free staggered spin susceptibility from expression (18), which is in analogy to the regularization procedure in calculating the chiral susceptibility \cite{Aoki, He}. That is to say, we define the regularized staggered spin susceptibility by
\begin{equation}
<S^{Z}(0)S^{Z}(0)>_R=<S^{Z}(0)S^{Z}(0)>-<S^{Z}(0)S^{Z}(0)>_{free},
\end{equation}
where the free staggered spin susceptibility $<S^{Z}(0)S^{Z}(0)>_{free}$
is calculated from Eq. (18) by setting $F(q^2)=1$, $A(p^{2})=1$ and 
$B(p^{2})=0$ \cite{a29}.
\begin{figure}[htbp]
\includegraphics[width=12cm]{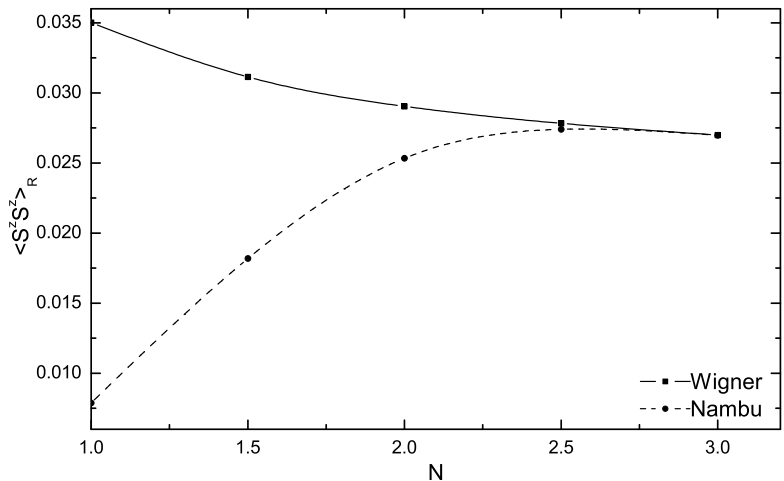}
\caption{The dependence of $<S^{Z}S^{Z}>_R$ on the number of fermion flavors in both Wigner phase and DCSB phase in the low energy limit}
\end{figure}

After solving the above coupled DSEs by means of
iteration method, we can now calculate the regularized staggered spin
susceptibility $<S^{Z}S^{Z}>_R$ given by Eq. (24) in the low energy limit in both Nambu phase and Wigner phase for the case of
$F(q^{2})=1$ and for several different number of fermion flavors. The dependence of $<S^{Z}S^{Z}>_R$ on the number of fermion flavors in both Wigner phase and DCSB phase is shown in Fig. 3. From Fig. 3 it can be seen that when $N$ approaches the critical number of fermion flavors, which is about 3.3 in the BC1 truncated scheme for DSE, the numerical values of the susceptibility in Nambu phase and Wigner phase are almost the same. This can be understood as follows. From Fig. 4 it can be seen that
when $N$ approaches the critical number of fermion flavors, the fermion propagator in Nambu phase tends to the one in Wigner phase. So the values of the susceptibility in these two phases should tend to be equal, which is what one expects in advance.
From Fig. 3 it can also be seen that for small $N$ the susceptibilities in these two phases show apparent difference, and 
here let us analyze the case of small $N$.
A physically interesting case is $N=2$. For $N=2$, the susceptibility in Wigner phase is 0.02904, and the susceptibility in Nambu phase is 0.02533.
The reason for this can be seen as follows. From Fig.1 
it can be seen that $A(p^2)$ in the two phases coincide in the large momentum region, but show apparent difference in the infrared region. 
In the infrared region $A(p^2)$ in Nambu phase is constant, whereas $A(p^2)$ in Wigner phase shows a power law behavior. From Fig. 2 it is also seen that $B(p^2)$ in Nambu phase is nearly a constant in the infrared region,
while it vanishes when $p^2$ is large enough. Therefore, it is easy to understand that when $N=2$, due to the difference between $A(p^2)$, $B(p^2)$ in these two phases in the infrared region, the numerical values of the susceptibility in these two phases are different. In addition, our numerical results show that
the smaller is $N$, the larger is the difference between the fermion propagators in Nambu
phase and Wigner phase (this can be seen by comparing Fig. 2 and Fig. 4). This explains why the smaller is $N$, the larger is the difference between the spin staggered susceptibility in Nambu phase and Wigner phase, as is shown in Fig. 3.

\begin{figure}[htbp]
\includegraphics[width=12cm]{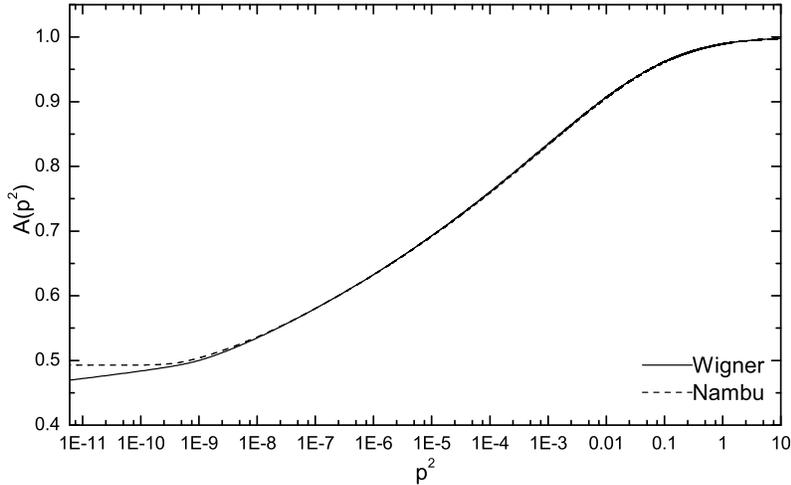}
\caption{The momentum dependence of $A(p^2)$ with $N=3$ in both Nambu phase and Wigner phase}
\end{figure}

In summary, in this paper, based on the linear response theory of
the fermion propagator to an external pseudoscalar field, we first obtain
a model-independent integral formula, which expresses the staggered spin susceptibility in terms of objects of the basic quantum field theory:  dressed propagator and vertex. When one approximates the pseudoscalar vertex function by the bare one, this expression, which includes the
influence of the nonperturbative dressing effects, reduces to the
expression for the staggered spin susceptibility obtained using
perturbation theory in previous works. After appropriately regularizing the additive linearly divergence, we study the staggered spin susceptibility by numerically solving the coupled DSEs in the low energy limit. Our results indicates that the
staggered spin susceptibility enhances and antiferromagnetic
correlation gets restored in both Nambu phase and Winger phase.

\acknowledgments

J.-F. Li thanks C.S. Fischer for discussion on the anomalous dimension.
This work is supported in part by the National Natural Science
Foundation of China (under Grant Nos 10935001, 11275097 and
11075075) and the Research Fund for the Doctoral Program of Higher
Education (under Grant No 2012009111002).

\end{document}